\documentstyle[12pt]{article}
\catcode`\@=11

\@addtoreset{equation}{section}
\def\theequation{\arabic{section}.\arabic{equation}}

\catcode`\@=11

\def\section{\@startsection{section}{1}{\z@}{3.5ex plus 1ex minus
   .2ex}{2.3ex plus .2ex}{\large\bf}}

\newskip\humongous \humongous=0pt plus 1000pt minus 1000pt

\newif\ifdtup

\def\eqnarray{\stepcounter{equation}\let\@currentlabel=\theequation
    \global\@eqnswtrue
    \global\@eqcnt\z@\tabskip\@centering\let\\=\@eqncr
    $$\halign to \displaywidth\bgroup\@eqnsel\hskip\@centering
      $\displaystyle\tabskip\z@{##}$&\global\@eqcnt\@ne
       \hfil${{}##{}}$\hfil
      &\global\@eqcnt\tw@ $\displaystyle\tabskip\z@{##}$\hfil
       \tabskip\@centering&\llap{##}\tabskip\z@\cr}
\def\lefteqn#1{\hbox to 4\arraycolsep{$\displaystyle #1$\hss}}
\def\thesection{\arabic{section}.}

\def\appendix{\setcounter{section}{0}
       \def\thesection{Appendix.}
       \def\theequation{\Alph{section}.\arabic{equation}}}

\long\def\@makefntext#1{\parindent 0cm\noindent
\hbox to 1em{\hss$^{\@thefnmark}$}#1}
\def\IR{{\hbox{{\rm I}\kern-.2em\hbox{\rm R}}}}
\def\IH{{\hbox{{\rm I}\kern-.2em\hbox{\rm H}}}}
\def\IC{{\ \hbox{{\rm I}\kern-.6em\hbox{\bf C}}}}
\def\IZ{{\hbox{{\rm Z}\kern-.4em\hbox{\rm Z}}}}
\newcommand{\beq}{\begin{equation}}
\newcommand{\eeq}{\end{equation}}
\def\rref#1{(\ref{#1})}
\newcommand{\ba}{\begin{array}{cc}}
\newcommand{\ea}{\end{array}}
\newcommand{\Lam}{|\Lambda|}
\newcommand{\M}{{\widetilde M}}
\newcommand{\Vol}{\hbox{\em Vol}\,}
\begin{document}

\begin{titlepage}
\vspace{.5in}
\begin{flushright}
UCD-92-8\\
May 1992\\
\end{flushright}
\vspace{.5in}
\begin{center}
{\LARGE\bf
Entropy vs.\ Action\\in the (2+1)-Dimensional\\
        Hartle-Hawking Wave Function}\\
\vspace{.4in}
{S.~C{\sc arlip}\footnote{\it email: carlip@dirac.ucdavis.edu}\\
       {\small\it Department of Physics}\\
       {\small\it University of California}\\
       {\small\it Davis, CA 95616}\\{\small\it USA}}
\end{center}

\vspace{.5in}
\begin{center}
{\large\bf Abstract}
\end{center}
{\small
In most attempts to compute the Hartle-Hawking ``wave function of the
universe'' in Euclidean quantum gravity, two important approximations
are made: the path integral is evaluated in a saddle point approximation,
and only the leading (least action) extremum is taken into account.
In (2+1)-dimensional gravity with a negative cosmological constant,
the second assumption is shown to lead to incorrect results: although
the leading extremum gives the most important single contribution to the
path integral, topologically inequivalent instantons with larger actions
occur in great enough numbers to predominate.  One can thus say that
in 2+1 dimensions --- and possibly in 3+1 dimensions as well --- entropy
dominates action in the gravitational path integral.
}
\end{titlepage}
\addtocounter{footnote}{-1}

\section{Introduction}

Quantum cosmology is a difficult subject, not least because we do not yet
have a consistent quantum theory of gravity.  In the absence of such a
theory, cosmologists must rely on plausible, but necessarily speculative,
approaches to gravity in the very early universe.  One attractive approach
is Hawking's Euclidean path integral \cite{Hawking}, which describes the
wave function of the universe in terms of a ``Wick rotated'' gravitational
path integral over Riemannian (positive definite) metrics $g$, with an
action
\beq
I_E[g,\phi] = -{1\over16\pi G}\int_M d^n\!x\, \sqrt{g}(R[g]-2\Lambda)
              -{1\over8\pi G}\int_{\partial M} d^{(n-1)}\!x\, \sqrt{h} K
              + I_{\rm matter}[\phi] \ .
\label{1x1}
\eeq
Here $R[g]$ is the scalar curvature for an $n$-dimensional manifold $M$
($n\!=\!4$ for standard physics), $\Lambda$ is the cosmological constant,
$h$ is the induced metric on $\partial M$, and $K$ is the trace of the
intrinsic curvature of $\partial M$, while $\phi$ represents a generic
set of matter fields.

A path integral ordinarily determines a transition amplitude between an
initial and a final configuration, and to specify a unique wave function
one must select appropriate initial conditions.  The Hartle-Hawking ``no
boundary'' proposal \cite{HartHawk} is that there should be {\em no}
initial geometry --- the path integral should be evaluated for compact
manifolds with only a single boundary component $\Sigma$.  If we
specify a metric $h$ and a set of matter fields $\phi|_\Sigma$ on
$\Sigma$, the path integral
\beq
\Psi[h,\phi|_\Sigma] = \int [dg][d\phi]\, \exp\left\{-I_E[g,\phi]\right\}
\label{1x2}
\eeq
can be interpreted as a wave function, giving an amplitude for the universe
to have spatial geometry $h$ and matter configuration $\phi|_\Sigma$.
There seems to be no natural way to select any one particular topology in
such a path integral, so in the Hartle-Hawking approach one sums over all
manifolds $M$, subject to the condition that $\Sigma$ be the sole boundary
component.

To obtain interesting physics,  a further restriction on $M$ must be
imposed.  The integration in \rref{1x2} is over Riemannian metrics, and it
is necessary to ``analytically continue'' to obtain the observed Lorentzian
structure of spacetime.  This will be possible if the Riemannian metrics
in the path integral can be joined to Lorentzian metrics to the future of
$\Sigma$ (see figure \ref{spacetime}). Gibbons and Hartle \cite{GibHart}
have shown that a finite action continuation across $\Sigma$ exists
only if the extrinsic curvature of $\Sigma$ vanishes.  One should
therefore limit the sum to manifolds $M$ and metrics $g$ such that the
boundary $\partial M = \Sigma$ is totally geodesic.

It is perhaps time to admit that no one knows how to evaluate a path
integral of the form \rref{1x2}.  Ordinary general relativity is
nonrenormalizable, so the standard perturbative expansions make no sense.
A standard procedure in quantum cosmology is to compute the path integral in
a saddle point approximation, in the hope that the resulting estimate may
be useful even if higher order corrections are not well-defined.  The result
is a kind of instanton picture, describing the semiclassical tunneling from
``nothing'' to a universe.  An extra implicit assumption is usually ---
although not always --- made: one considers only the extrema with the
smallest actions,\footnote{In wormhole sums \cite{Coleman,GidStrom}, one
includes a class of higher action metrics as well --- essentially an
instanton gas of the leading extrema --- but most of the classical
solutions are still neglected.} which are assumed to dominate the path
integral.

It would clearly be useful to have a well-defined model in which these
assumptions could be tested more carefully.  Gravity in 2+1 dimensions
provides such a model.  It is by now fairly well known that pure
(2+1)-dimensional gravity is renormalizable \cite{Witten,Deser}, and that
a canonically quantized theory can be formulated with no approximations
\cite{Witten,Moncrief,Car:time}.  At the same time, many of the basic
features of (3+1)-dimensional gravity --- the form of the action,
diffeomorphism invariance, the possibility of summing over topologies ---
remain unaltered.

In a series of very interesting papers, Fujiwara et al.\
\cite{Fuji1}--\cite{Fuji3} have begun to investigate quantum cosmology for
pure (2+1)-dimensional gravity with a negative cosmological constant.  They
have found the leading saddle point contribution to the path integral, and
have investigated the possibilities of topology change and the spontaneous
creation of particle-like defects in spacetime.  The purpose of this paper
is to extend this work by examining the contributions of higher action
extrema.  We shall see that the classical solutions discussed by Fujiwara
et al.\ do not, in fact, determine the overall structure of the
Hartle-Hawking wave function; instead, the wave function is peaked at
certain highly symmetric two-geometries for which an infinite number of
distinct topologies contribute to the path integral.

\section{The Path Integral in 2+1 Dimensions}

Let us begin with a brief discussion of the path integral \rref{1x2} in
2+1 dimensions with a negative cosmological constant and no matter.  The
classical equations of motion coming from the action \rref{1x1} are
\beq
R_{ik} = -2\Lam g_{ik} \ .
\label{2x2}
\eeq
In three dimensions, the Ricci tensor completely determines the curvature,
and \rref{2x2} implies that $g_{ik}$ is a constant negative curvature
metric, i.e., that $(M,g)$ is a hyperbolic three-manifold.  The
Hartle-Hawking wave function depends on the boundary value $h$, which is
itself a hyperbolic metric on the totally geodesic boundary $\Sigma$. This
is in accord with the canonically quantized theory \cite{Witten,Moncrief,HN},
where wave functions are quite naturally expressed as square integrable
functions on the moduli space ${\cal M}_\Sigma$ of hyperbolic metrics on
$\Sigma$.

On general grounds, we expect the contribution of an extremum $(M,\bar g)$
to the path integral to take the form
\beq
\Psi_M\!\left[{\bar g}|_{\lower1pt\hbox{$\scriptstyle\Sigma$}}\right]
    = \Delta_M e^{-I_E[\bar g]} \ ,
\label{2x1}
\eeq
where $I_E[\bar g]$ is the classical action at the extremum and the
prefactor $\Delta_M$ is a combination of determinants coming from small
fluctuations around $\bar g$ and from gauge-fixing.  Using \rref{2x2}, we
see that the exponent is
\beq
I_E[\bar g] = {1\over4\pi G\Lam^{1/2}}\Vol(M) \ ,
\label{2x3}
\eeq
where $\Vol(M)$ is the volume of the three-manifold $M$ with the metric
rescaled to constant curvature $-1$.  This volume is a well-known
topological invariant in three-manifold theory: a given manifold will
admit at most one hyperbolic metric, and it can be shown that at most a
finite number of three-manifolds have any given hyperbolic volume
\cite{Gromov}.

To evaluate the prefactor $\Delta_M$, we can appeal to the relationship
between the three-dimensional Einstein action and the Chern-Simons action
for the gauge group PSL($2,\!\IC$).  As Witten has observed
\cite{Witten:Jones}, the first order form of the Einstein action in three
dimensions with $\Lambda<0$ can be rewritten as
\beq
I_{CS} = {1\over 64\pi i G\Lam^{1/2}} \int_M d^3\!x\, \epsilon^{ijk}\!
  \left[ A_i^a(\partial_j A_{ka} - \partial_k A_{ja})
  + {2\over3}\epsilon_{abc}A_i^aA_j^bA_k^c\right] + \hbox{c.c.} \ ,
\label{2x4}
\eeq
where the complex gauge field
\beq
A_i^a
  = {\scriptstyle{1\over2}}\epsilon^{abc}\omega_{ibc} + i\Lam^{1/2}e_i^a
\label{2x5}
\eeq
is a PSL($2,\!\IC$) connection, expressed in terms of a ``dreibein'' or
frame field $e_i^a$ and a spin connection $\omega_{i\,b}^{\,a}$ on $M$.
The extrema of \rref{2x4} are flat PSL($2,\!\IC$) connections, and it is
easy to check that the condition of flatness is equivalent to the field
equations \rref{2x2} for the metric $g_{ik} = e_i^a e_{ka}$.  At the same
time, \rref{2x4} may also be recognized as the standard Chern-Simons action
for PSL($2,\!\IC$), allowing us to use known results for Chern-Simons path
integrals to evaluate $\Delta_M$.

To understand this correspondence better, it is helpful to know a bit more
about the geometric significance of the connection $A_i^a$.  Recall first
that PSL($2,\!\IC$) is the isometry group of hyperbolic three-space $\IH^3$.
Any three-manifold $M$ with a constant negative curvature metric is locally
isometric to $\IH^3$, and can be covered by coordinate charts $U_\alpha$
isometric to $\IH^3$ with transition functions $\phi_{\alpha\beta}$ in
PSL($2,\!\IC$).  This ``geometric structure'' allows us to define a natural
flat $s\ell(2,\!\IC)$ bundle $E$ over $M$ as follows \cite{Hodgson}: we
first construct the product bundle $s\ell(2,\!\IC)\times U_\alpha$ on each
chart, and then identify the fibers in the overlap $U_\alpha\cap U_\beta$
by means of the adjoint action of the transition function
$\phi_{\alpha\beta}$.  It can then be shown that the connection $A_i^a$
is precisely the flat connection on $E$.

Equivalently, at least if $M$ is geodesically complete, we can write
\beq
M = \IH^3/\Gamma
\label{2x6}
\eeq
for some discrete subgroup $\Gamma\subset \hbox{PSL}(2,\!\IC)$, unique up
to conjugacy class.  $\Gamma$ is called the holonomy group of $M$; it is
isomorphic to the fundamental group $\pi_1(M)$, and can be viewed as a
representation of $\pi_1(M)$ in PSL($2,\!\IC$).   The bundle $E$ is then
\beq
E = \left(s\ell(2,\!\IC)\times\IH^3\right)/\Gamma \ ,
\label{2x7}
\eeq
where the quotient is by the simultaneous action of $\Gamma$ as a group
of isometries of $\IH^3$ and the adjoint action of $\Gamma$ on
$s\ell(2,\!\IC)$.  This description makes it clear that $E$ is completely
determined by the group $\Gamma$.

Standard results from Chern-Simons theory now tell us that
\beq
\Delta_M = T^{1/2}(M,E) \ ,
\label{2x8}
\eeq
where $T(M,E)$ is the Ray-Singer torsion, or equivalently\footnote{For
noncompact groups, this equivalence is discussed in \cite{W-BN}.
Schwarz and Tyupkin \cite{Schwarz} have pointed out that an additional
anomaly can occur because $M$ has a boundary.  This problem will not
arise when the boundary is totally geodesic, however, since one can
then compute everything on the (closed) double of $M$.} the
Reidemeister-Franz torsion, associated with the flat bundle $E$.
Strictly speaking, we must use a slight modification of the standard
definition of Ray-Singer torsion, as discussed in reference \cite{W-BN},
because PSL($2,\!\IC$) is noncompact.  For a complex gauge group, the
effect of this change is essentially to replace torsion computed in
terms of $A$ alone with its absolute square; heuristically, the partition
function receives separate contributions from the path integrals over $A$
and $\bar A$ in the action \rref{2x4}.  To obtain \rref{2x8} from
reference \cite{W-BN}, we have also used the fact that the connection
$A_i^a$ is isolated (by rigidity theorems for hyperbolic structures
\cite{Mostow}) and irreducible, and that the framing anomaly and the
phase of the prefactor $\Delta_M$ both vanish for PSL($2,\!\IC$),
essentially because of a cancellation between left- and right-moving
modes \cite{W-BN,Witten:cpx}.

The Chern-Simons formulation can potentially give us information about
higher order corrections to the saddle point approximation as well.  In
particular, although the higher order terms have not been computed
explicitly, we know that they will be of order $G\Lam^{1/2}$, and will
thus be small if the cosmological constant is sufficiently small
\cite{Witten:Jones}.

Combining \rref{2x3} and \rref{2x8}, we obtain a contribution to the wave
function of the form
\beq
\Psi_M[h] = T^{1/2}(M,E)\,
 \exp\left\{ -{\Vol(M)\over\,4\pi G\Lam^{1/2}}\right\} \
\label{2x9}
\eeq
for each extremum $(M,\bar g)$ with $\bar g|_\Sigma = h$.  These
contributions must be summed over extrema,
\beq
\Psi[h] = \!\sum_{M\in{\cal I}(\Sigma,h)}\!\Psi_M[h]\ ,
\label{2x10}
\eeq
where the ``space of instantons'' ${\cal I}(M,h)$ comprises all hyperbolic
manifolds with induced hyperbolic metric $h$ on a single totally geodesic
boundary $\Sigma$.  Our next task is to categorize this space.

\section{Counting Hyperbolic Manifolds}

The classification of hyperbolic three-manifolds is one of the most
active areas of research in modern topology, and for now we should not
expect to completely understand the space of extrema of the
(2+1)-dimensional Einstein action.  We may still look for particular
points in ${\cal I}(M,h)$, however, that can give us useful information
about the Hartle-Hawking wave function \rref{2x10}.

Let us first note that for most hyperbolic metrics $h$ on $\Sigma$, there
are {\em no} saddle point contributions --- the field equations \rref{2x2}
usually have no solution with a specified boundary value for the metric.
It is known, however, that solutions exist for a dense set of values of
$h$ in the moduli space ${\cal M}_\Sigma$ \cite{Thurston-Fujii}.  If
the full path integral behaves reasonably smoothly, the saddle point
approximation for $\Psi[h]$ on such a dense set should be adequate for
physics.

Kojima and Miyamoto \cite{KoMi} have recently found the hyperbolic
manifolds of least volume with a single totally geodesic boundary of
any given genus.  These are the extrema $M_R$ considered by Fujiwara et
al.\ \cite{Fuji1}.  For a given spatial topology $\Sigma$, the
corresponding manifold $M_R(\Sigma)$ will have some definite boundary
metric $h_R$, and since the wave function \rref{2x9} is exponentially
suppressed for large volumes, we might expect $\Psi[h]$ to be peaked at
the corresponding two-geometry.

On the other hand, a given extremum $M_R$ makes only a single contribution
to $\Psi[h_R]$.  We must also ask whether other spatial metrics $h$ are
boundary values for large numbers of extrema of the action.  If this is
the case, the number of instantons --- the ``entropy'' --- may overcome
the exponential volume suppression.

To see that this is possible, we consider a family of hyperbolic
three-manifolds discovered by Neumann and Reid \cite{NR1,NR2}.  (The
manifolds most relevant to physics are actually not quite the ones
described in these references, but rather a closely related family found
by Alan Reid; see the appendix for details.)  The family consists of an
infinite number of manifolds $\M_{(p,q)}$, where $p$ and $q$ are relatively
prime integers, with the following characteristics:
\begin{enumerate}
\addtolength{\itemsep}{-4pt}
\item each of the $\M_{(p,q)}$ has a single totally geodesic boundary,
with a fixed hyperbolic metric $h_\infty$ that is independent of $p$ and
$q$;
\item the volumes of the $\M_{(p,q)}$ are bounded above by a finite number
$\Vol(\M_\infty)$, and converge to $\Vol(\M_\infty)$ as
$p^2+q^2\rightarrow\infty$; and
\item the Ray-Singer torsions $T(\M_{(p,q)},E_{(p,q)})$ considered in
the previous section do not converge as $p^2+q^2\rightarrow\infty$, but
instead take on a dense set of values in the interval $(0,cT_\infty]$,
where $cT_\infty$ is a positive constant.
\end{enumerate}

These properties imply that the $\M_{(p,q)}$ all give positive
contributions to the Hartle-Hawking wave function at $h\!=\!h_\infty$.
Indeed, conditions (2) and (3) guarantee that the sum over topologies
diverges: the volumes converge to $\Vol(\M_\infty)$, while
infinitely many of the prefactors are bounded below by some $\epsilon>0$.
The Hartle-Hawking wave function is thus infinitely peaked at $h_\infty$.

The construction of the families $\M_{(p,q)}$ is discussed in more detail
in the appendix, but the basic procedure is fairly easy to describe.
Neumann and Reid start with a finite volume hyperbolic orbifold $M_\infty$
(note no tilde here) with two essential characteristics: a totally geodesic
boundary consisting of a sphere with three conical singularities, and
a cusp $K$ that is separated from this boundary.  The singularity at $K$
can be ``filled in'' by a standard procedure called hyperbolic Dehn
surgery --- essentially by replacing a neighborhood of the cusp with
a solid torus --- to obtain a set of new orbifolds $M_{(p,q)}$, where the
integers $(p,q)$ describe the way the torus is twisted before it is glued
in.  This surgery procedure cannot change the hyperbolic structure on
the boundary, however, since a sphere with three cone points admits
only one hyperbolic metric.

Neumann and Reid then consider a covering space $\M_\infty$ of $M_\infty$
in which the orbifold singularities, including the conical singularities
on the boundary, are ``unwrapped.''  The cusp $K$ of $M_\infty$ is lifted
to a  set of cusps $\widetilde K$ on $\M_\infty$,
and Dehn surgeries on $M_\infty$ lift to surgeries on $\M_\infty$.
These lifted surgeries must again leave the boundary of $\M_\infty$
fixed, since otherwise their projections to $M_\infty$ would change the
boundary there.  We thus obtain an infinite family of nonsingular
three-manifolds all having an identical totally geodesic boundary.

It can be shown that Dehn surgery on a cusp of a hyperbolic manifold
always decreases the volume \cite{Thurston}.  This fact gives us condition
(2) above: the volume of $\M_{(p,q)}$ is bounded above by the volume of the
cusped manifold $\M_\infty$.  For $p^2+q^2$ large, Neumann and Zagier
\cite{NZ} have found a fairly simple description of the rate of convergence
of the volume $\Vol(\M_{(p,q)})$ to $\Vol(\M_\infty)$:
\beq
\# \left\{(p,q): \Vol(M_{(p,q)})<\Vol(M_\infty) - 1/x\right\}
  = 6\pi x + O(x^{1/2}) \ .
\label{3x1}
\eeq

To obtain condition (3), we must understand the behavior of the
Ray-Singer torsion under hyperbolic Dehn surgery.  Our basic strategy
parallels Witten's computation of Chern-Simons amplitudes on surgered
manifolds \cite{Witten:Jones}: we separately compute the torsions
of $\M_{(p,q)}\!-\!V$ and $V$, where $V$ is the solid torus added by the
surgery.  We can then use the ``gluing theorem'' for torsion \cite{Vishik}
to obtain
\beq
T(\M_{(p,q)},E_{(p,q)}) = T(\M_{(p,q)}\!-\!V,E_{(p,q)}|_{\M-V})\cdot
                         T(V,E_{(p,q)}|_V) \ .
\label{3x2}
\eeq
By construction, $\M_{(p,q)}\!-\!V$ is diffeomorphic to $\M_\infty$; its
torsion differs from that of $\M_\infty$ only because the hyperbolic
structure --- the flat bundle $E_{(p,q)}$ --- differs.  But for $p^2 + q^2$
large, the holonomy groups $\Gamma(\M_{(p,q)}\!-\!V)$ converge to
$\Gamma(\M_\infty)$, which is sufficient to show that the torsions
converge.  For $V$, on the other hand, the Ray-Singer torsion can be
calculated explicitly (see appendix for details).  The result again
depends on the flat bundle, and thus on $p$ and $q$; one finds that
\beq
T(V,E_{(p,q)}|_V) = {c\over4}(\cosh 2\ell - \cos 2t)^2 \ ,
\label{3x3}
\eeq
where $\ell$ and $t$ are the length and torsion of the core geodesic of
$V$. (``Torsion'' here means not the Ray-Singer torsion, but the ordinary
geometric torsion of the geodesic as a curve in three-space.)  For $p$
and $q$ large, $\ell$ converges to zero, but it is known \cite{Meyerhoff}
that $t$ takes on a dense set of values in the interval $[0,2\pi]$, so
$T(V,E_{(p,q)})$ varies rapidly in the range $(0,c]$.  Our result then
follows directly from inserting \rref{3x3} into \rref{3x2}.

It would be interesting to find a more detailed description of the
behavior of the parameter $t$ for $p^2+q^2$ large, ideally leading to
a result for the Ray-Singer torsion analogous to \rref{3x1}.  Such a
description would allow us to approximate the sum over $p$ and $q$ by
an integral, perhaps permitting a more quantitative description of
the divergence of the wave function.

\section{Implications}

We have seen that the Hartle-Hawking wave function \rref{2x10} diverges
for at least one value of the spatial metric $h$.  A natural question is
how often this behavior occurs.  If $\Psi[h]$ diverges for most values of
$h$, our result is essentially negative --- we will have merely shown that
the ``leading instanton'' approximation is not valid.  If such divergences
are relatively rare, on the other hand, we may have learned a good deal
about $\Psi[h]$.

Let us first consider the exponent in \rref{2x9}. The volume $\Vol(M)$
may be viewed as a real-valued function on the space of instantons
${\cal I}(M,h)$.  In our example, we produced a bounded sequence of
volumes by hyperbolic Dehn surgery on a cusped manifold.  For closed
hyperbolic three-manifolds, it can be shown that this is the only way to
produce such a sequence; in particular, the only accumulation points of
the volume in $\IR$ correspond to manifolds with at least one cusp
\cite{Gromov,Thurston}.

By a simple doubling argument, the same is true for manifolds with a
totally geodesic boundary.  Hence the kind of divergence we saw above
will only occur for values of $h$ that can be realized as boundary values
of hyperbolic metrics on cusped manifolds.  Most hyperbolic manifolds have
no cusps, of course, so this is likely to be a significant restriction,
although as far as I know this issue has not been investigated by
topologists.

The Neumann-Reid construction suggests --- although it does not prove ---
a much stronger restriction.  In general, one expects hyperbolic
Dehn surgery to change the boundary metric of a three-manifold, smearing
out any divergence in the sum over topologies.  This did not occur in our
example for a very specific reason: the boundary we have been considering
can be realized as a covering space of a rigid surface, the two-sphere with
three conical singularities.  (``Rigid'' means that the surface admits
only one constant negative curvature metric, i.e., that its moduli space
is a single point.)  Only a few, highly symmetric surfaces occur as
covering spaces of rigid surfaces, and it is plausible that the sum over
topologies will diverge only for such surfaces.  If this is the case, it
may be possible to give a complete description of the normalized
Hartle-Hawking wave function as a sum of delta functions at isolated
points in moduli space.

The key question, of course, is whether such results can be extended
to 3+1 dimensions.  For $n=4$, the extrema of the action \rref{1x1} need
not have constant negative curvature, and a detailed analysis becomes
much more difficult.  But we can at least ask how the constant negative
curvature manifolds contribute to the wave function.

Four-dimensional hyperbolic manifolds have a
discrete set of volumes, with only finitely many manifolds having the
same volume.  In contrast to the three-dimensional case, there are no
longer any accumulation points.  It is still true, however, that the
number of manifolds of a given volume can grow rapidly as the volume
increases.  According to Gromov \cite{Gromov}, the number of hyperbolic
four-manifolds with volume less than $x$ may grow as fast as
\beq
x\exp(\exp(\exp(4+x))) \ .
\label{4x1}
\eeq
If the increase is nearly this rapid, we may again expect entropy to
dominate action in the Hartle-Hawking wave function of the universe.

Moreover, it is plausible that the sum over topologies will again be
dominated by highly symmetric spacetimes.  Topologically distinct
hyperbolic manifolds with the same volume typically arise when the
boundaries of a fundamental polyhedron can be glued together to form a
manifold in more than one way.  As in the (2+1)-dimensional case, this
may be viewed as an indication of underlying symmetry.  This connection
is admittedly speculative, however; a more quantitative description
would clearly be of interest.

\vspace{2.5ex}
\begin{flushleft}
\large\bf Acknowledgements
\end{flushleft}

This work would not have been possible without help from a number of
mathematicians.  I would especially like to thank Alan Reid for his
explanations of families of hyperbolic manifolds, and Joel Hass for
teaching me some needed hyperbolic geometry.  I am grateful for additional
help from Colin Adams, Walter Carlip, Craig Hodgson, Geoff Mess, Walter
Neumann, Mel Rothenberg, Albert Schwarz, and Jeff Weeks.  I would also
like to thank the readers and distributors of the USENET newsgroup
sci.math.research for providing a medium through which I was able to
contact several people who helped me with this work.  This research was
supported in part by the Department of Energy under grant DE-FG03-91ER40674.

\newpage
\appendix
\section{Mathematical Details}
\setcounter{footnote}{0}

The purpose of this appendix is to fill in some of the mathematical
details omitted from the text.  We will discuss hyperbolic Dehn surgery,
the Neumann-Reid construction, and the computation of Ray-Singer torsion.

\subsection{Hyperbolic Dehn Surgery}

In this section, we briefly summarize hyperbolic Dehn surgery
\cite{Thurston,Hodgson}.  A cusp of a hyperbolic three-manifold $M$ can be
viewed as an embedded circle that is ``infinitely far away'' from the rest
of the manifold in the hyperbolic metric.  Topologically, a neighborhood
of a cusp is diffeomorphic to $T^2\times[t_0,\infty)$, where $T^2$ is a
two-dimensional torus. (The circle itself is not part of $M$, which is
complete but not compact.)  Metrically, we can take the upper half space
model for $\IH^3$, with the standard constant negative curvature metric
\beq
ds^2 = t^{-2}(dx^2 + dy^2 + dt^2) \ ;
\label{A1}
\eeq
a neighborhood of a cusp then looks like a region
\beq
N = \left\{ (x,y,t): t>t_0, z\sim z+1, z\sim z+\tau\right\}
\label{A2}
\eeq
(see figure \ref{cusp}).  Note that for fixed $t$, the metric \rref{3x2}
is Euclidean, so the constant $t$ cross-sections of \rref{A2} are ordinary
flat tori with modulus $\tau$.

To perform (topological) Dehn surgery, we remove a small neighborhood $K$
of a cusp --- or more generally, of any embedded circle in $M$ --- and
replace it with a solid torus $V = S^1\times D^2$.  $M\!-\!K$ has a toroidal
boundary where $K$ has been cut out, and associated with this boundary are
two commuting generators of $\pi_1(M\!-\!K)$, say $m$ and $\ell$.  To glue
in the solid torus $V$, we first choose a closed curve $\gamma$ on the
boundary along which to attach a cross-section $\{p\}\!\times\!D^2$ of $V$.
Once this disk is glued in, there is no remaining topological freedom,
since $V\!-\!\{p\}\!\times\!D^2$ is topologically simply a ball, which
can be glued in uniquely.

The curve $\gamma$ can be written in terms of the generators
$\ell$ and $m$ as
\beq
\gamma = m^p\ell^q \ .
\label{A3}
\eeq
This expression will represent a simple closed curve if $p$ and $q$ are
relatively prime integers.  The effect of surgery on the fundamental group
is to add one relation $m^p\ell^q=1$ to $\pi_1(M\!-\!K)$, that is, to kill
one generator.

In general, one can say little about the geometry of a manifold resulting
from surgery.  In fact, {\em any} three-manifold can be obtained from Dehn
surgery on a link in the three-sphere \cite{Lickorish}.  However, Thurston
\cite{Thurston} has shown that if one performs Dehn surgery on a cusp of a
hyperbolic three-manifold, the resulting manifold will itself admit a
hyperbolic metric for all but a finite number of choices of $p$ and $q$.
The process of creating this new hyperbolic manifold is called hyperbolic
Dehn surgery.

Thurston's result can be restated as follows.  A cusped manifold admits a
unique {\em complete} hyperbolic metric.  But it also admits an infinite
number of incomplete hyperbolic metrics, parameterized by relatively prime
integers $p$ and $q$, that can be completed to give nonsingular, complete
hyperbolic metrics by adding an appropriate solid torus.  This formulation
makes it easy to describe one sense in which the manifolds $M_{(p,q)}$
converge to $M_\infty$ as $p^2+q^2\rightarrow\infty$: it can be shown
that the corresponding distance functions converge \cite{Thurston,Milnor}.
Pictorially, a cusp may be visualized as an infinitely long, exponentially
shrinking tube with a toroidal cross-section (figure \ref{cusp}); as $p$
and $q$ become large, the surgery affects the manifold farther and farther
out on this tube, leaving an increasingly large piece of $M_\infty$
essentially unchanged.

This convergence is also reflected algebraically in the holonomy groups
$\Gamma_{(p,q)}$.  An element of $\Gamma_{(p,q)}$ represents a closed
geodesic in $M_{(p,q)}$, and as $p^2+q^2\rightarrow\infty$, the geodesics
converge to the geodesics of $M_\infty$.  Consequently, the holonomy
groups also converge; in particular, if $g_{(p,q)}\in\Gamma_{(p,q)}$
represents a curve $\gamma$ in $M_{(p,q)}\!-\!K$, then the $g_{(p,q)}$ will
converge to the element $g\in\Gamma_\infty$ representing $\gamma$ in the
holonomy of $M_\infty$.  This result will be important below in the
analysis of Ray-Singer torsion.

\subsection{The Neumann-Reid Construction}

The fundamental construction used in this paper is based on the papers
\cite{NR1} and \cite{NR2} of Neumann and Reid.  The particular variation
used here is unpublished, and was explained to me by Alan Reid.

An orbifold is a space locally modelled on $\IR^n/\Gamma$, where $\Gamma$
is a finite group that acts properly discontinuously but not necessarily
freely.  The set of points of $\IR^n$ at which the action is not free
projects down to a set called the singular locus of the orbifold.  A
typical two-dimensional orbifold is a cone of order $n$, ${\cal O} =
\IR^2/C_n$, where $C_n$ is the group generated by rotations by $2\pi/n$
around some point $p$.  $C_n$ acts freely except at $p$, and the singular
locus is thus the apex of the cone.   $C_n$ also acts on $\IR^3$ by
rotation around an axis; the singular locus is then a line corresponding
to this axis.  By a (three-dimensional) hyperbolic orbifold, we mean an
orbifold locally modelled on $\IH^3/\Gamma$, where $\Gamma$ is now a group
of isometries of hyperbolic three-space.  Orbifold singularities then come
from torsion elements of $\Gamma$.

In \cite{NR1}, Neumann and Reid construct a set of hyperbolic three-orbifolds
$M_\infty{(m,n)}$, each having an underlying space $S^2\times[0,\infty)$
and a boundary consisting of a totally geodesic two-sphere $\Sigma{(m,n)}$
with three conical points.  The singular locus of one of these orbifolds
is shown in figure \ref{orbifold}.  A line in this diagram labeled by
an integer $n$ corresponds to a cone angle of $2\pi/n$; a vertex at which
lines labeled $2$, $2$, and $n$ meet is locally $\IR^3/D_{n}$, where
$D_n$ is the dihedral group of order $2n$.

The bottom boundary of figure \ref{orbifold} is thus a sphere with cone
points of cone angles $2\pi/2$, $2\pi/m$, and $2\pi/n$.  The top boundary,
on the other hand, is a ``pillow cusp,'' with a neighborhood of the form
$K\approx F\times[t_0,\infty)$, where $F$ is a sphere with four cone points
of cone angle $\pi$.  $F$ has Euler characteristic zero, and can be expressed
as the quotient of a torus by the cyclic group of order two; that is, a
pillow cusp has an ordinary cusp as a double cover.

We previously defined hyperbolic Dehn surgery for an ordinary cusp, but
it can be shown that a similar procedure is possible for a pillow cusp.
As in the case of an ordinary cusp, surgery on the cusp of $M_\infty{(m,n)}$
produces a family of orbifolds $M_{(p,q)}{(m,n)}$, all but a finite number
of which admit hyperbolic structures.  Moreover, this surgery cannot affect
the boundary $\Sigma{(m,n)}$, since a sphere with three conical points
is rigid, that is, it admits a unique hyperbolic metric.

Now suppose that for some $m$ and $n$, we can find a new manifold
$\M_\infty{(m,n)}$ that is a finite covering space of $M_\infty{(m,n)}$,
such that the boundary $\Sigma(m,n)$ lifts to a single connected surface
$\widetilde\Sigma(m,n)$.  In such a covering, the cusp $K$ will lift to a
set of cusps $\widetilde K$, and $(p,q)$ surgery on $K$  will lift to some
$(p',q')$ surgery on $\widetilde K$.  The boundary $\widetilde\Sigma(m,n)$
must be left invariant by such $(p',q')$ surgery, since it projects back
to the fixed boundary $\Sigma(m,n)$ of $M_\infty{(m,n)}$.  The set of such
$(p',q')$ surgeries on $\widetilde K$ --- that is, surgeries that are
equivariant with respect to the cover --- is an infinite subset of the
set of all surgeries, and we will have thus obtained an infinite family of
manifolds with identical totally geodesic boundary $\widetilde\Sigma(m,n)$.

We must thus show that such a covering can exist.  To do so, let us choose
$m=3$ and $n>6$ a prime number such that $n=1\ (\hbox{mod}\ 4)$.  We shall
need a presentation of the orbifold fundamental group of $M_\infty{(3,n)}$:
\beq
\Gamma(3,n)= \langle a, b, c, d~|~a^2 = b^2 = c^3 = d^n
                     = (ab)^2 = (bc)^2 = (cd)^2 = (da)^2 = 1\rangle \ ,
\label{AB1}
\eeq
where the subgroup representing the fundamental group of the boundary
$\Sigma(3,n)$ is the triangle group
\beq
\Delta(2,3,n) = \langle c, d~|~c^3 = d^n = (cd)^2 =1 \rangle \ .
\label{AB2}
\eeq
This presentation can be obtained by applying standard techniques
\cite{Wielen} to the construction of reference \cite{NR1}, in which
$M_\infty(m,n)$ is described explicitly in terms of reflections in the
faces of a polyhedron.  Alternatively, it can be read off from figure
\ref{orbifold}; a loop around a line with cone angle $2\pi/r$ represents
an element in $\Gamma(m,n)$ of order $r$, while relations come from the
requirement that the product of loops completely encircling a vertex be
contractible.

We will have completed our proof if the following conditions are satisfied:
\begin{enumerate}
\addtolength{\itemsep}{-4pt}
\item there exists an epimorphism $\phi$ from $\Gamma(3,n)$ to a finite
group $G_n$ such that $\Delta(2,3,n)$ surjects onto $G$; and
\item the kernel $\kappa$ of $\phi$ is torsion free.
\end{enumerate}
To see that these conditions are sufficient, observe that the cover
corresponding to the subgroup $\kappa\subset\Gamma$ is a finite cover of
$M_\infty(3,n)$ with fundamental group $\kappa$; by condition (2), this
covering space is a manifold, with no orbifold singularities.  Then by
standard covering space theory, condition $(1)$ implies that the preimage
of $\Sigma{(3,n)}$ is a connected surface (see, for instance, section 5.11
of reference \cite{Massey}).

To complete the proof, we must therefore construct the map $\phi$ to a
finite group.  Let $G_n$ be the finite simple group $PSL(2,{\bf F}_n)$,
where ${\bf F}_n$ is the field of $n$ elements.  For $n$ prime such that
$n=1\ (\hbox{mod}\ 4)$, some elementary number theory shows that $-1$ is
a square in ${\bf F}_n$.  Let $x\in{\bf F}_n$ be such that $x^2=-1$, and
let $z=-1+2x$, $t = -4x(1+2x)^{-1}$.  We then define the epimorphism $\phi$
from $\Gamma(3,n)$ to $PSL(2,{\bf F}_n)$ by
\beq
a \mapsto \left(\ba x&t\\0&-x\ea\right) ,\quad
b \mapsto \left(\ba 2&z\\2\!+\!z&-2\ea\right) ,\quad
c \mapsto \left(\ba 0&1\\-1&1\ea\right) ,\quad
d \mapsto \left(\ba 1&1\\0&1\ea\right) \ .
\label{AB3}
\eeq
Note that these each have determinant $1$ in $PSL(2,{\bf F}_n)$.  We then
have
\beq
ab \mapsto \left(\ba -2x&xz\!-\!2t\\2\!-\!x&2x\ea\right) ,\quad
bc \mapsto \left(\ba -z&2\!+\!z\\2&z\ea\right) ,\quad
cd \mapsto \left(\ba 0&1\\-1&0\ea\right) ,\quad
da \mapsto \left(\ba x&t\!-\!x\\0&-x\ea\right) \ ,
\label{AB4}
\eeq
and it is easily checked that the relations in \rref{AB1} are all satisfied.
Moreover, $PSL(2,{\bf F}_n)$ is generated by $({1\atop 0}\,{1\atop 1})$ and
$({{\ 0}\atop -1}\,{1\atop 1})$, so $\Delta(2,3,n)$ surjects onto $G_n$.
Finally, as every element of finite order is conjugate in $\Gamma(3,n)$
to one of the elements \rref{AB3}, \rref{AB4}, the kernel is torsion-free
\cite{Wielen}.  Our two requirements are thus met, and the proof is
completed.

For these examples, we can compute the genus of the boundary
$\widetilde\Sigma(3,n)$ explicitly.  The Euler characteristic of
$\Sigma(3,n)$ is $\chi(\Sigma(3,n)) = (6-n)/6n$ \cite{Thurston}, while
by construction, the degree of the covering is equal to the order of
$PSL(2,{\bf F}_n)$.  Thus
\beq
2-2g = \chi(\widetilde\Sigma(3,n)) = {6-n\over 6n}|PSL(2,{\bf F}_n)|
  = {(6-n)(n^2-1)\over 12} \ .
\label{AB5}
\eeq
It may checked that the resulting genus is always an integer --- although
typically a rather large one --- when $n=1\ (\hbox{mod}\ 4)$.

\subsection{Computing Torsions}

In this section, we compute Ray-Singer torsions for surgered manifolds.
(See \cite{Schwarz,Ray-Singer} for more detailed definitions of these
torsions.)  Our starting point is a ``gluing formula.''  Let $E$ be a
flat bundle over a manifold $M$, and suppose that $M$ can be written as
a union of two pieces $M_1$ and $M_2$ joined along a common boundary
$\Sigma$. Vishik \cite{Vishik} has then shown that
\beq
T(M,E) = T(M_1,E|_{M_1})\cdot T(M_2,E|_{M_2})\cdot T(\Sigma,E|_\Sigma) \ ,
\label{AC1}
\eeq
where the determinants on $M_1$ and $M_2$ are defined with relative
(Dirichlet) boundary conditions.  In particular, for hyperbolic $(p,q)$
Dehn surgery on a cusp of $M_\infty$, we can take $M_1$ to be the solid
torus $V$ added by surgery, $M_2$ to be $M_\infty\!-\!K$, and $E$ to be
the flat $s\ell(2,\!\IC)$ bundle (see section 2) corresponding to the new
hyperbolic structure on the surgered manifold $M_{(p,q)}$.  The common
boundary $\Sigma$ is then a two-dimensional torus, and the last term in
\rref{AC1} can be omitted, since the torsion of a closed even-dimensional
manifold is always trivial.

We must first compute the torsion $T(V,E_{(p,q)})$ for $V =
S^1\!\times\!D^2$.  Since $D^2$ is simply connected, we can use the
product formula\footnote{This relationship is proven by Ray and Singer
for closed manifolds \cite{Ray-Singer}, but the extension to manifolds
with boundary with relative boundary conditions is straightforward.}
\beq
T(S^1\!\times\!D^2,E) = T(S^1,E|_{S^1})^{\chi(D^2)} \ ,
\label{AC2}
\eeq
where $\chi$ is the Euler number, $\chi(D^2)=1$.  We thus need merely
evaluate the Ray-Singer torsion for a circle.  As Bar-Natan and Witten
have noted \cite{W-BN}, there are some subtleties involved in the
definition of the Laplacian because $s\ell(2,\!\IC)$ is noncompact, but
it is not hard to check that the appropriate Ray-Singer torsion is
\beq
T(S^1,E) = |{\det}^\prime\Delta_0| \ ,
\label{AC3}
\eeq
where the Laplacian acts on $s\ell(2,\!\IC)$-valued functions $\phi
= \phi^a t_a$ twisted by the holonomy $H$,
\beq
\phi(\theta+2\pi) = H^{-1}\phi(\theta)H \ .
\label{AC4}
\eeq
Up to an overall conjugation that does not affect the determinants, the
holonomy around $S^1$ takes the form
\beq
H = \left( \begin{array}{cc}
            e^{\ell+it} & 0\\
            0 & e^{-(\ell+it)}
            \end{array} \right) \ ,
\label{AC5}
\eeq
where $\ell$ and $t$ have a geometrical interpretation as the length and
torsion of the ``core geodesic'' $S^1$ of $V$.  The evaluation of the
determinant \rref{AC3} is then reasonably straightforward: eigenfunctions
take the form
\beq
\phi(\theta) = \left( \begin{array}{cc}
              a(\theta) & b(\theta)\\
              c(\theta) & -a(\theta)
              \end{array} \right)
\label{AC6}
\eeq
with
\begin{eqnarray}
a(\theta+2\pi)&=&a(\theta) \nonumber \\
b(\theta+2\pi)&=&e^{-2(\ell+it)}b(\theta) \\
c(\theta+2\pi)&=&e^{2(\ell+it)}c(\theta) \nonumber \ ,
\label{AC7}
\end{eqnarray}
and $\zeta$-function regularization gives
\beq
T(S^1,E) = 16\pi^2(\cosh 2\ell - \cos 2t)^2 \ .
\label{AC8}
\eeq

An important caveat is necessary here: equation \rref{AC8} is not quite
independent of the metric in $S^1$.  The Laplacian \rref{AC3} has
zero-modes --- constant matrices $\phi$ that commute with the holonomy
--- and the Ray-Singer torsion should really be viewed as a section of
the line bundle $(\det H^0)^{-1}(\det H^1)$.  A canonical section exists
only if one has a metric with which to normalize harmonic forms; the
expression \rref{AC8} was computed with respect to the metric $d\theta^2$
on the circle.  This metric dependence must drop out of the final expression
for the torsion of the surgered manifold $M_{(p,q)}$, since the Laplacians
there have no zero modes (see below), but it would be interesting to
understand how this happens in more detail.  It was because of this
ambiguity that the constant $c$ was left unspecified in equation \rref{3x3}.

It is instructive to recompute \rref{AC8} in terms of Reidemeister-Franz
torsion (see \cite{Lott} or \cite{Milnor:torsion} for a clear explanation
of this invariant; note, however, that the quantity $\tau$ in reference
\cite{Lott} is equal to $-\log T$ in our conventions).  One begins with
the cellular version of the flat bundle \rref{2x7},
\beq
{\cal C}(\widetilde V, \partial\widetilde V; E) =  s\ell(2,\!\IC)
  \otimes_{\lower3pt\hbox{$\scriptstyle\Gamma$}}
  {\cal C}(\widetilde V,\partial\widetilde V) \ ,
\label{AC9}
\eeq
where ${\cal C}(\widetilde V,\partial\widetilde V)$ is a (relative) chain
complex for the universal covering space $\widetilde V$ of $V$, and the
holonomy group $\Gamma\!=\!\langle H \rangle \approx \pi_1(V)$ acts by
deck transformations on $\widetilde V$ and by the adjoint action on
$s\ell(2,\!\IC)$.  We can take the Pauli matrices $\sigma^a$ and $i\sigma^a$
as generators for $s\ell(2,\!\IC)$, and it is evident from figure \ref{cell}
that a basis for ${\cal C}(\widetilde V, \partial\widetilde V; E)$ consists
of the three-cells $\{\sigma^a\otimes S, (i\sigma^a)\otimes S\}$ and the
two-cells $\{\sigma^a\otimes F, (i\sigma^a)\otimes F\}$, with a boundary
operator
\beq
\partial (\sigma^a\otimes S)
         = \left\{ \begin{array}{ll}
           (H^{-1}\sigma^a H-\sigma^a)\otimes F & \mbox{if $a=1,2$}\\
            0 & \mbox{if $a=3$}
            \end{array} \right.
\label{AC10}
\eeq
(with a similar expression for the cells involving $i\sigma^a$).  The
relative homology group $H_3$ is clearly generated by $\{\sigma^3\otimes S,
(i\sigma^3)\otimes S\}$, while $H_2$ is generated by $\{\sigma^3\otimes F,
(i\sigma^3)\otimes F\}$.

To compute Reidemeister-Franz torsion, we need volume forms for $C_q$
and $H_q$.  The volume forms for the chain groups are determined by the
preferred basis fixed by the cell decomposition of $V$,
\begin{eqnarray}
\omega_2 &=& \bigwedge \left(\sigma^a\otimes F\right)\bigwedge
             \left((i\sigma^a)\otimes F\right) \nonumber \\
\omega_3 &=& \bigwedge \left(\sigma^a\otimes S\right)\bigwedge
             \left((i\sigma^a)\otimes S\right)  \ ,
\label{AC11}
\end{eqnarray}
while volume forms for the homologies take the form
\begin{eqnarray}
\mu_2 &=& h \left(\sigma^3\otimes F\right) \wedge
            \left((i\sigma^3)\otimes F\right) \nonumber \\
\mu_3 &=& k \left(\sigma^3\otimes S\right) \wedge
            \left((i\sigma^3)\otimes S\right)
\label{AC12}
\end{eqnarray}
for some constants $h$ and $k$.  We now choose an arbitrary volume form
for $B_2 = \partial C_3$, say
\beq
\rho = \partial\,(\sigma^1\otimes S)\wedge \partial\,(\sigma^2\otimes S)
\wedge\partial\,((i\sigma^1)\otimes S)\wedge \partial\,((i\sigma^2)\otimes S)
       \ ,
\label{AC13}
\eeq
and write
\begin{eqnarray}
\omega_2 &=& m_{\hbox{\scriptsize even}}\, \rho\wedge \mu_2 \nonumber \\
\omega_3 &=& m_{\hbox{\scriptsize odd}}\, (\partial^{-1}\rho)\wedge \mu_3
      \ .
\label{AC14}
\end{eqnarray}
The Reidemeister-Franz torsion is then defined to be
\beq
T({\cal C}(\widetilde V, \partial\widetilde V; E)) =
   m_{\hbox{\scriptsize odd}}/m_{\hbox{\scriptsize even}}
  \ .
\label{AC15}
\eeq

It is easy to check that this expression agrees with \rref{AC8}, up to a
constant factor depending on $h$ and $k$.  This ambiguity again reflects
the existence nontrivial homology; the two expressions will agree
completely if we define the volume forms $\mu_2$ and $\mu_3$ on the
homology to be dual to normalized volume forms on the cohomology, where
the normalization once again depends on the choice of metric.  It is
worth noting that the dependence of the torsion on the holonomy group
$\langle H \rangle$ comes entirely through the boundary operator
\rref{AC10}, and can be expressed in terms of the determinant of the
``combinatorial Laplacian'' $\partial^\dagger\partial +
\partial\partial^\dagger$.

We are now left with the second term in \rref{AC1}, the torsion of the
manifold $M_2 = M_\infty\!-\!K$ with flat bundle $E_{(p,q)}$.  This
quantity is rather difficult to compute in general, but some conclusions
about its behavior can be reached.  Note first that the topology of $M_2$
is independent of the choice of the surgery coefficients $p$ and $q$;
in fact, $M_2$ is diffeomorphic to the cusped manifold $M_\infty$.  Let
us choose a cell decomposition for this manifold once and for all, with
a corresponding set of deck transformations $\gamma_i\in\pi_1(M_\infty)$
that determine the gluing pattern of the cells.  The Reidemeister-Franz
torsion will again be computed from a finite set of determinants, whose
entries are fixed by the cell decomposition and by the representation
$\Gamma_{(p,q)}$ of $\pi_1(M_2)$ in PSL($2,\!\IC$).  In particular, the
only dependence of the torsion on the surgery coefficients will come
through the dependence of the combinatorial Laplacian on a set of elements
$g_{(p,q)}(\gamma_i)\in\Gamma_{(p,q)}$ that represent the gluing maps for
$M_{(p,q)}\!-\!K$.

But we saw above that the representations $\Gamma_{(p,q)}$ converge to
$\Gamma_\infty$ as $p^2+q^2\rightarrow \infty$.  In particular, the
elements $g_{(p,q)}(\gamma_i)$ converge.  Hence the determinants of the
combinatorial Laplacians must also converge, and the torsions
$T(M_\infty\!-\!K,E_{(p,q)})$ will converge to some number $T_\infty$,
which can be interpreted as the Reidemeister-Franz torsion for the
original cusped manifold $M_\infty$ (with relative boundary conditions
at the cusp).  This guarantees that $T_\infty$ is nonvanishing, since
Reidemeister-Franz torsion is always nonzero.

Finally, let us return to the question of zero-modes and the possible
metric dependence of the total torsion $T(M_{(p,q)},E_{(p,q)})$.  In
principle, this torsion is again defined as a section of a line bundle,
\beq
(\det H^0)^{-1}(\det H^1)(\det H^2)^{-1}(\det H^3) \approx
(\det H^0)^{-2}(\det H^1)^2 \ .
\label{AC16}
\eeq
In the case of interest to us, however, this bundle does not arise.
Elements of $H^0(M_{(p,q)}; E_{(p,q)})$ represent global Killing vectors
of $M_{(p,q)}$ \cite{Hodgson}, which are certainly generically absent.
Elements of $H^1(M_{(p,q)};E_{(p,q)})$ represent infinitesimal variations
of the connection $A^a_i$ in the space of flat connections, that is,
infinitesimal deformation of the hyperbolic structure.  But the boundary
of $M_{(p,q)}$ is totally geodesic, so $M$ can be replaced by its (closed)
double for the purpose of analyzing these deformations.  The Mostow rigidity
theorem \cite{Mostow} then guarantees that no such deformations exist.
Thus $H^0$ and $H^1$ are trivial, and any metric dependence must ultimately
drop out of the torsion \rref{AC1}.

\renewcommand{\textfraction}{0}
\renewcommand{\topfraction}{1}
\newpage

\begin{figure}[b]
\caption{A spacetime that will contribute to the Hartle-Hawking wave
function.  The metric is Riemannian to the past of $\Sigma$ and
Lorentzian to the future.}
\label{spacetime}
\end{figure}
\begin{figure}[b]
\caption{The neighborhood of a cusp of a hyperbolic manifold, represented
in the upper half space model of $\IH^3$.  Note that the area of a toroidal
cross section is proportional to $t^{-2}$, while the distance from $t_0$
to $t$ is $d = \ln(t/t_0)$, so the area decreases exponentially with proper
distance.}
\label{cusp}
\end{figure}
\begin{figure}[b]
\caption{The orbifold that forms the starting point of the Neumann-Reid
construction.  A line labeled by an integer $n$ represents a singularity
with cone angle $2\pi/n$.  The underlying space of this orbifold is
$S^2\times[0,\infty)$; horizontal cross-sections of the diagram should
be interpreted as two-spheres.}
\label{orbifold}
\end{figure}
\begin{figure}[b]
\caption{A cell decomposition for the universal covering space $\widetilde V$
of the solid torus $V$.  Relative to the boundary $\partial\widetilde V$, the
only cells are the three-cell $S$, the two-cell $F$, and their translates by
deck transformations.}
\label{cell}
\vspace{-.75in}
\end{figure}

\vspace{4ex}



\begin{thebibliography}{99}
\bibitem{Hawking} S.\ W.\ Hawking, in {\em General Relativity: An Einstein
   Centenary Survey}, S.\ W.\ Hawking and W.\ Israel, editors (Cambridge
   University Press, 1979).
\bibitem{HartHawk} J.\ B.\ Hartle and S.\ W.\ Hawking, {\em Phys.\ Rev.}
   {\bf D28}, 2960 (1983).
\bibitem{GibHart} G.\ W.\ Gibbons and J.\ B.\ Hartle, {\em Phys.\ Rev.}
   {\bf D42}, 2458 (1990).
\bibitem{Coleman} S.\ Coleman, {\em Nucl.\ Phys.} {\bf B310}, 643 (1988).
\bibitem{GidStrom} S.\ B.\ Giddings and A.\ Strominger, {\em Nucl.\ Phys.}
   {\bf B307}, 854 (1988).
\bibitem{Witten} E.\ Witten, {\em Nucl.\ Phys.} {\bf B311}, 46 (1988).
\bibitem{Deser} S.\ Deser, J.\ McCarthy, and Z.\ Yang, {\em Phys.\ Lett.}
   {\bf B222}, 61 (1989).
\bibitem{Moncrief} V.\ Moncrief, {\em J.\ Math.\ Phys.} {\bf 30}, 2907
   (1989).
\bibitem{Car:time} S.\ Carlip, {\em Phys.\ Rev.} {\bf D42}, 2647 (1990).
\bibitem{Fuji1} Y.\ Fujiwara et al., {\em Phys.\ Rev.} {\bf D44}, 1756
   (1991).
\bibitem{Fuji2} Y.\ Fujiwara et al., {\em Phys.\ Rev.} {\bf D44}, 1763
   (1991).
\bibitem{Fuji3} Y.\ Fujiwara et al., ``Point Particles as Defects in
   (2+1)-Dimensional Quantum Gravity, Tokyo Inst.\ Tech.\ preprint
   TIT-HEP-171 (1991).
\bibitem{HN} A.\ Hosoya and K.\ Nakao, {\em Class.\ Quantum Grav.} {\bf 7},
   163 (1990).
\bibitem{Gromov} M.\ Gromov, in {\em S\'eminaire Bourbaki 1979/80,
   Expos\'es 543-560} (Lecture Notes in Mathematics {\bf 842}), A.\ Dold
   and B.\ Eckmann, editors (Springer, 1981).
\bibitem{Witten:Jones} E.\ Witten, {\em Commun.\ Math.\ Phys.} {\bf 121},
   351 (1989).
\bibitem{Hodgson} C.\ Hodgson, ``Degeneration and Regeneration of Geometric
   Structures on Three-Manifolds,'' Princeton University dissertation (1986).
\bibitem{W-BN} D.\ Bar-Natan and E.\ Witten, {\em Commun.\ Math.\ Phys.}
   {\bf 141}, 423 (1991).
\bibitem{Schwarz} A.\ S.\ Schwarz and Yu.\ S.\ Tyupkin, {\em Nucl.\ Phys.}
   {\bf B242}, 436 (1984).
\bibitem{Mostow} G.\ D.\ Mostow, {\em Ann. of Math.\ Studies} {\bf 78}
   (Princeton University Press, 1973).
\bibitem{Witten:cpx} E.\ Witten, {\em Commun.\ Math.\ Phys.} {\bf 137},
   29 (1991).
\bibitem{Thurston-Fujii} W.\ P.\ Thurston, cited in M.\ Fujii, {\em Osaka
   J.\ Math.} {\bf 27}, 539 (1990).
\bibitem{KoMi} S.\ Kojima and Y.\ Miyamoto, {em J.\ Differential Geometry}
   {\bf 34}, 175 (1991).
\bibitem{NR1} W.\ D.\ Neumann and A.\ W.\ Reid, {\em Math.\ Proc.\ Camb.\
   Phil.\ Soc.}~{\bf 109}, 509 (1991).
\bibitem{NR2} W.\ D.\ Neumann and A.\ W.\ Reid, ``Rigidity of Cusps in
   Deformations of Hyperbolic 3-Orbifolds,'' Ohio State University preprint
   (1991).
\bibitem{NZ} W.\ D.\ Neumann and D.\ Zagier, {\em Topology} {\bf 24}, 307
   (1985).
\bibitem{Vishik} S.\ M.\ Vishik, {\em Sov.\ Math.\ Dokl.} {\bf 36}, 174
   (1988).
\bibitem{Meyerhoff} R.\ Meyerhoff, in {\em Low Dimensional Topology and
   Kleinian Groups} (London Math.\ Society Lecture Notes Series {\bf 112}),
   D.\ B.\ A.\ Epstein, editor (Cambridge University Press, 1986).
\bibitem{Thurston} W.\ P.\ Thurston, ``The Geometry and Topology of
   Three-Manifolds,'' Princeton lecture notes (1979).
\bibitem{Lickorish} W.\ B.\ R.\ Lickorish, {\em Ann.\ Math.} {\bf 76},
   531 (1962).
\bibitem{Wielen} A.\ M.\ Bruuner et al., {\em Topology and its Applications}
   {\bf 20}, 289 (1985).
\bibitem{Massey} W.\ S.\ Massey, {\em Algebraic Topology: An Introduction}
   (Springer, 1977).
\bibitem{Milnor} J.\ Milnor, {\em Bull.\ Amer.\ Math.\ Soc.} {\bf 6}, 9
   (1982).
\bibitem{Ray-Singer} D.\ B.\ Ray and I.\ M.\ Singer, {\em Adv.\ Math.}
   {\bf 7}, 145 (1971).
\bibitem{Lott} J.\ Lott, in {\em Recent Developments in Geometry}, S.-Y.\
   Cheng, H.\ Choi, and R.\ E.\ Greene, editors (American Math.\ Society,
   1989).
\bibitem{Milnor:torsion} J.\ Milnor, {\em Bull.\ Amer.\ Math.\ Soc.}
   {\bf 173}, 358 (1966).


\end{thebibliography}
\end{document}